\documentclass[preprint,prd,floatfix,nofootinbib,superscriptaddress,showpacs]{revtex4}
 \usepackage{graphicx}
\def\Z0{${\em Z^0\/}$}

\def\r#1 {$^{#1}$}

\newcommand{\gevc} { {\rm GeV/c}}
\newcommand{\gevcc}{ {\rm GeV/c^2}}

\def\gepsfcentered#1{
  \def\testit{#1}
  \def\lbracket{[}
  \ifx\testit\lbracket
    \let\dofilecmd=\gepsfwithopt
  \else
    \let\dofilecmd=\gepsfnoopt
  \fi
  \dofilecmd}

\def\gepsfnoopt#1{
  \begin{center}
  \leavevmode
  \epsffile{#1}
  \end{center}}

\def\gepsfwithopt#1 #2 #3 #4]#5{
  \begin{center}
  \leavevmode
  \gepsfmaxx=0.94\textwidth
  \epsffile[#1 #2 #3 #4]{#5}
  \end{center}}

\newdimen\gepsfmaxx
\def\epsfsize#1#2{
  \ifnum \epsfxsize=0
    \ifnum \epsfysize=0
      \ifnum #1 > \gepsfmaxx
        \gepsfmaxx
      \else
        #1
      \fi
    \else
      \epsfxsize
    \fi
  \else
    \epsfxsize
  \fi
}

\begin{document}
 \bibliographystyle{unsrt}
 \title {Phenomenological study of the atypical heavy flavor
         production observed at the Fermilab Tevatron}
%
%
 \vskip 0.025in
 \affiliation{Laboratori Nazionali di Frascati, Istituto Nazionale 
              di Fisica Nucleare, Frascati, Italy\\}
 \affiliation{Fermi National Accelerator Laboratory, Batavia, 
              Illinois 60510, USA\\}
 \affiliation{Istituto Nazionale di Fisica Nucleare, I-56100 Pisa, Italy\\}
 \affiliation{University of Cyprus, 1678 Nicosia, Cyprus\\}
 \vspace{0.2em}
 \author{G.~Apollinari}
 \affiliation{Fermi National Accelerator Laboratory, Batavia, 
              Illinois 60510, USA\\}
 \author{M.~Barone}
 \affiliation{Laboratori Nazionali di Frascati, Istituto Nazionale 
              di Fisica Nucleare, Frascati, Italy\\}
 \author{I.~Fiori}
 \affiliation{Istituto Nazionale di Fisica Nucleare, I-56100 Pisa, Italy\\}
 \author{P.~Giromini}
 \author{F.~Happacher}
 \author{S.~Miscetti}
 \author{A.~Parri}
 \affiliation{Laboratori Nazionali di Frascati, Istituto Nazionale 
              di Fisica Nucleare, Frascati, Italy\\}
 \author{F.~Ptohos}
 \affiliation{University of Cyprus, 1678 Nicosia, Cyprus\\}
 \vspace{0.2em}
 \begin{abstract}
   We address known discrepancies between the heavy flavor properties of jets 
   produced at the Tevatron collider and the prediction of conventional-QCD 
   simulations. In this study, we entertain the possibility that these effects are
 real and due to new physics. We
   show that all anomalies can be simultaneously fitted by postulating 
   the additional pair production of light bottom squarks with a 100\% 
   semileptonic branching fraction. \\
 \end{abstract} 
 \preprint{FERMILAB-PUB-05-375-E}
 \maketitle

 \section {Introduction} \label{sec:ss-intro} 
  In 2001 the CDF experiment reported~\cite{suj} the observation of an 
  excess of events - containing a lepton with large transverse momentum, large
  transverse missing energy, and two or three jets one of which (called
  superjet) contains a displaced secondary vertex and a soft lepton due to the
  decay of presumed $b$ hadrons - over the standard model (SM) expectation   
  (17 such events are observed and $4.9\pm 0.6$ are expected)~\footnote{
  The D${\not\! {\rm O}}$ collaboration has also searched~\cite{d0_suj} for the
  presence of such an anomaly, and reports a deficit with respect to the SM 
  prediction (2 events observed and $3 \pm 0.5$  predicted). 
  Unfortunately, because of
  stricter selection requirements and a soft lepton tagging efficiency
  significantly smaller than that of CDF, the resulting 95\% C.L. limit to the cross section
  for producing this type of events is at least a factor of two larger than the
  cross section corresponding to the excess observed by CDF~\cite{suj-int}.}.

  The  apparent excess of superjets observed by CDF
  has been modeled in  Ref.~\cite{suj-int} by
  postulating the unconventional production of a low-mass, strongly 
  interacting object that decays semileptonically with a branching fraction of
  the order of 1 and a lifetime of the order of the picosecond. Since at the
  time there were no limits to the existence of a charge $-1/3$ scalar quark
  lighter than 10 $\gevcc$~\cite{kane,nappi,carena}, the supersymmetric
  partner of the bottom quark was a potential candidate. The hypothesis of
  a very light scalar quark has been  investigated in many theoretical
  papers and has prompted careful re-examinations of precision electroweak 
  data and $e^+\;e^- \rightarrow$ hadrons data~\cite{squa-1,squa-2,squa-3,
  squa-4,squa-5,squa-6,squa-7,squa-8,squa-9,squa-10,squa-11,squa-12,squa-13,
  squa-14,squa-15}. It is a fair summary of all this work that no data 
  analysis shows convincing evidence for a very light scalar quark. 
  The conjecture itself is not favored, but also not completely excluded,
  in the framework of the minimal supersymmetric standard model (MSSM).
  The CLEO collaboration~\cite{cleo-sb} has searched for pair production of
  light squarks that decay to a lepton, a charmed quark and a scalar neutrino
  with a 100\% branching fraction by  selecting events that contain two
  oppositely charged leptons plus a $D$ meson below the $B\bar{B}$ threshold.
  Reference~\cite{cleo-sb} reports no evidence for the production of such
  squarks with mass between 3.5 and $4.5\; \gevcc$. However, an exclusion
  limit depends on the
  modeling of the scalar quark hadronization  and  decay matrix element, 
  and  squarks lighter than 3.8 $\gevcc$ could have escaped
  detection~\cite{cleo-sb}.  
 
  Because of the unresolved experimental situation, the CDF collaboration has
  performed a detailed comparison~\cite{ajets} between measured and predicted
  heavy-flavor properties of jets produced at the Tevatron, intended to search
  for evidence either supporting or disfavoring this conjecture. 
  Reference~\cite{ajets} studies rates of observed and predicted leptons 
  arising from $b$-quark decays in jets 
 that recoil against a generic jet or recoil against a jet that also
 contains a lepton. That study
 finds that in the second case the rate of jets containing a lepton from
 presumed $b$ decays is 50\% higher than in the first case. The magnitude of
 the effect is consistent with what is expected in the presence of
  pair production of light squarks with a $100$\% semileptonic branching ratio. 
  The magnitude  of the discrepancy is  consistent with the result of
  other
  measurements performed at the Tevatron.
  The ratio of the $\mu+\bar{b}$-jet cross section
  to the next-to-leading-order (NLO) 
  prediction~\footnote{
  As discussed in Ref.~\cite{ajets}, the NLO prediction for producing two
  central bottom quarks above the same transverse momentum threshold has an
  uncertainty no larger than 15\%; in contrast, the NLO prediction of the
  single $b$ cross section has a $\simeq 50$\% uncertainty.} is 
   measured to be 
  1.5$\pm$10\%~\cite{derw}, the ratio of the cross
  section for producing dimuons from presumed $b$- and $\bar{b}$-hadron semileptonic decays
  to the NLO prediction is  2.6$\pm$20\%~\cite{yale,d0-yale}, whereas this
  ratio is 1.1$\pm$20\% when identifying $b$ quarks by locating secondary vertices
  produced by their decays~\cite{ajets,tara,bbar}.
  If one assumes that the above mentioned measurements and their quoted uncertainties 
  are correct, the disagreement between data and theory appears to be
  a  function of the number of semileptonic decays used to
  identify $b$ quarks~\footnote{This effect could also be a fortuitous result
  produced by the
 fact that  the distribution of the ratios of the $b\bar{b}$
  cross section measurements at the Tevatron, with and without leptons,
 to  NLO prediction has approximately a $ 30$\% RMS deviation, 
 considerably larger than the quoted
 experimental uncertainties~\cite{bbar}.}.
  Reference~\cite{sameside} has extended the comparison
  of Ref.~\cite{ajets} to jets containing a lepton pair. In that study, the
  distributions of
  the dilepton invariant mass and opening angle are found to be
  strikingly different
  from those predicted by a conventional-QCD simulation, in which most lepton
  pairs arise from sequential semileptonic decays of single $b$ hadrons.
  In addition, a study of the invariant mass distribution of muon pairs collected
  by the CDF experiment~\cite{dimu} has attempted to rule out $1^{--}$
  bound states of light scalar quarks that are not excluded by the SPEAR 
  searches for narrow resonances~\cite{nappi,spear}. Reference~\cite{dimu}
  improves the SPEAR limit by almost an order of magnitude, and sets an
  average upper credible limit of approximately $8$ eV to the leptonic width
  of new narrow resonances below the $\Upsilon$ mesons. An exception is the
  mass of $7.2\; \gevcc$ at which an excess of $250 \pm 61$ events over a
  smooth background is observed. The statistical significance of the signal 
  ($3.5\; \sigma$) is not enough to establish the existence of a new particle.
  However, the size of this signal is consistent with the production of a
  $1^{--}$ bound state of spin~0 and charge~$-1/3$ quarks.

  All anomalies listed above certainly require further and independent
  confirmation.
  However, we entertain the possibility that all these
  effects are real, and verify if a new physics process 
 can simultaneously model 
  the discrepancies between data and conventional-QCD predictions reported in
  Refs.~\cite{ajets,sameside}. Since we use the same data sets, simulated
  samples, and analysis tools we briefly
  recall those studies.
  Reference~\cite{ajets} compares a QCD simulation based upon the 
  {\sc herwig}~\cite{herwig} and {\sc qq}~\cite{cleo} Monte Carlo generator 
  programs to a data sample, referred to as inclusive lepton sample, that
  consists of events with two or more jets, one of which is required to
  contain a lepton with $p_T \geq 8\; \gevc$ (trigger lepton); this jet is
  referred to as lepton-jet, whereas jets recoiling against it are called 
  away-jets. That study uses measured rates of lepton- and away-jets 
  containing displaced secondary vertices (SECVTX tags) or tracks with large
  impact parameter~\cite{jpb} (JPB tags) in order to determine the bottom
  and charmed content of the data. In order to remove the uncertainty of the
  predicted heavy flavor production cross sections, the simulation is tuned
  to match the heavy-flavor content of the data. In Ref.~\cite{ajets} rates
  of SECVTX and JPB tags, as well as the event kinematics, are well modeled
  after tuning the parton-level cross sections predicted by the {\sc herwig}
  generator within the experimental and theoretical uncertainties. However,
  the data have a $50$\% excess of events in which one additional jet 
  (away-jet) contains a soft lepton with $p_T \geq 2\; \gevc$ (SLT tag). 
  Reference~\cite{sameside} has extended the comparison between data and
  tuned simulation to lepton-jets that, in addition to the trigger lepton, 
  contain an additional soft lepton with opposite sign charge (OS). After
  removing the background arising from fake leptons by subtracting
  rates of lepton pairs with same sign charge (SS), the yields of OS-SS 
  lepton pairs are compared to the tuned simulation. For invariant masses 
  smaller than $2\; \gevcc$, the data are found to be largely underestimated
  by the simulation.

  In the present study we try to model the same data used in
  Refs.~\cite{ajets,sameside} by implementing the conventiona-QCD simulation, referred 
  to as SM simulation in the following, with the
  hypothetical pair production of sbottom 
  quarks ( $b_s$).
  Section~\ref{sec:ss-scaqua} describes the implementation of this new
  process into the SM simulation. In Sec.~\ref{sec:ss-fit} we tune the SM+$b_s$
  simulation to also reproduce measured yields of jets with SLT and dilepton
  tags. Section~\ref{sec:ss-inv} investigates the sensitivity of the data to
  other models. Section~\ref{sec:ss-scalifet} presents a crude estimate of the
  lifetime of the hypothetical object causing the excess of SLT tags.
  Our conclusions are presented in Sec.~\ref{sec:concl}.  
 \section{Production and decay of sbottom quarks} \label{sec:ss-scaqua}
  The production of $b_s$ quarks that decay to $l\; X$ with a 100\% branching
  ratio would be sufficient to model the excess of away-jets with SLT tags
  reported in Ref.~\cite{ajets}. In order to model the excess of jets
  containing a lepton pair reported in Ref.~\cite{sameside}, we need an extra
  source of leptons. This could be provided by the sequential
  semileptonic decay of $c$ quarks, and we choose
  the decay $b_s \rightarrow c \; l\; \nu_s$,
  where $\nu_s$ is a massless non-interacting scalar.

\begin{figure}
 \begin{center}
 \leavevmode
 \includegraphics*[width=\textwidth]{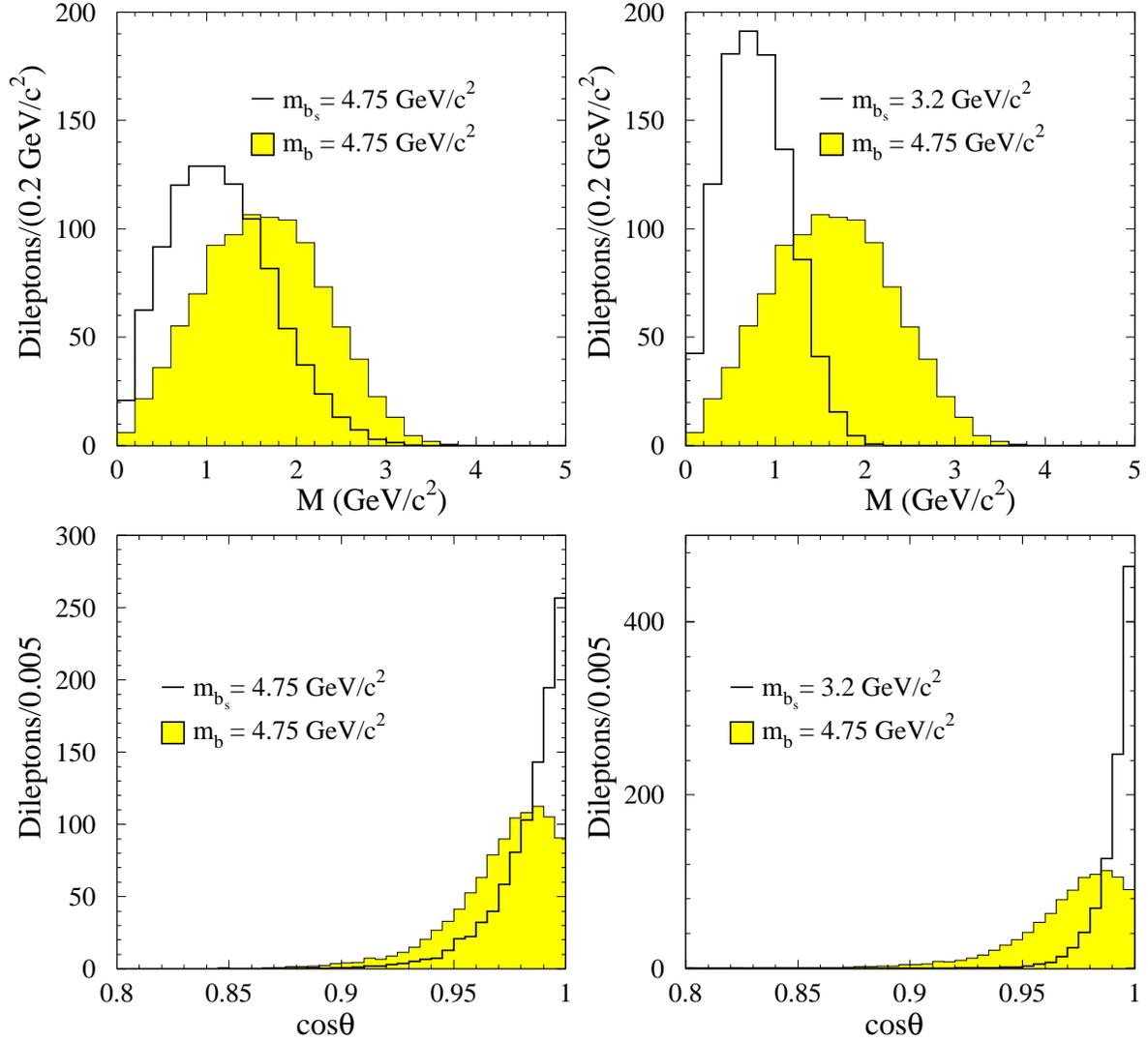}
 \caption[]{Distributions of the invariant mass $M$ and the opening angle
            $\theta$ between  leptons arising from  the sequential decays
            $b \rightarrow l\;  c \;\nu$, with $ c \rightarrow l\; s \; \nu$
            and $b_s \rightarrow l\; c\; \nu_{s} $, with 
            $c \rightarrow l\: s\; \nu $. We set $m_c=1.5\; \gevcc$.
            Leptons are not simulated through the detector.}
 \label{fig:fig_scaqua_1}
 \end{center}
 \end{figure}

  We model the $b_s$ decay with the matrix element discussed in more detail in 
  Appendix~\ref{sec:ss-matr}. In this model, the decay is mediated by the
  higgsino coupling to the right-handed matter.
  This matrix element produces lepton pairs from sequential 
  decay that have an invariant mass distribution different from that  
  produced by the  $V-A$ matrix element of conventional heavy-flavor decays
  (see Fig.~\ref{fig:fig_scaqua_1}). We implement the production of pairs
  of scalar quarks in $p\bar{p}$ collisions into the {\sc herwig} generator
  with the method described in Appendix~\ref{sec:ss-mod}. The method starts
  with the generation of $b\bar{b}$ pairs with a quark mass of $3.65\; \gevcc$
  and a lifetime of 1 ps. At the end of the $b$-hadronization process, the
  formed $B$ hadrons are turned into fictitious $\tilde{B}$ hadrons. We have
  generated samples of simulated events with the same luminosity of the
  conventional-QCD samples used in Refs.~\cite{ajets,sameside}. The NLO
  calculation of the process $p \bar{p} \rightarrow b_s \bar{b}_s$,
  implemented into the {\sc prospino} Monte Carlo generator~\cite{prosp},
  predicts a cross section that is approximately $15$\% of the NLO cross
  section for producing pairs of quarks with the same mass~\cite{mnr}.
  Therefore the parton-level 
  $b_s$  cross section in the {\sc herwig} simulated samples is overestimated
  by a factor of seven.
\newpage
 \section{Fit of the data with the SM simulation implemented with pair
          production of scalar quarks} \label{sec:ss-fit}
  In Ref.~\cite{ajets}, after tuning the SM simulation to fit the
  observed rates of lepton- and away-jets with SECVTX and JPB tags due to
  heavy flavor, we have observed a discrepancy between measured and predicted
  rates of away-jets with soft lepton tags. In this study, we use a simulation
  that includes pair production of $b_s$ quarks with $100$\% semileptonic
  branching ratio; we tune this production by also fitting rates of away-jets
  with SLT tags. Rates of tagged jets in the data are taken from Tables~II
  and~VII of Ref.~\cite{ajets}. We also fit rates of (OS-SS) lepton pairs 
  ($Dil$) contained in the same jet, taken from Table~III of 
  Ref.~\cite{sameside}. We use the same notations and fitting procedure
  described in Sec.~VIII of Ref.~\cite{ajets}.

  In the fit we tune the parton-level cross section of the different
  heavy-flavor production mechanisms [direct production (LO), flavor
  excitation, and gluon splitting] to reproduce the rates of tags observed
  in the data. Following the procedure of Ref.~\cite{ajets},
  we use two fit parameters to account 
  for the uncertainty of the luminosity and selection cuts of the electron and
  muon data; 
  these factors include the uncertainty of the $b$ direct production.
  Five fit parameters are used to determine the relative weight of the 
  remaining parton-level cross sections for producing conventional heavy 
  flavors ($b$ and $c$) with respect to the $b$ direct production.
  These five parameters are constrained to the default values
  as described in Sec.~VIII of Ref.~\cite{ajets}. Direct production
  and flavor excitation of  $b_S$ quarks are weighted with two additional
  fit parameters. As explained in Appendix~\ref{sec:ss-mod}, the value of
  the flavor excitation cross section returned by the fit also includes
  gluon splitting contributions. The efficiency for
  finding SECVTX and JPB tags in the simulation have been tuned using a 
  subset of the inclusive lepton sample~\cite{cdf-tsig}. Since we are now
  entertaining the hypothesis that this data sample might be contaminated
  by unconventional production, in the fit we also weight the $b$-
  and $c$-quark tagging efficiencies
  with free parameters (scale factors)
  to allow for the possibility that they might be wrong.
  We assume that the simulated efficiency for tagging $b_s$ hadrons is
  correct~\footnote{
  We have arbitrarily fixed the $b_s$ lifetime; the introduction of a scale
  factor for the simulated $b_s$ tagging efficiency is equivalent to 
  introducing an uncertainty on the lifetime. The dependence on the lifetime
  is studied in Sec.~\ref{sec:ss-scalifet}.}.
 
  The result returned by the best fit with the SM+$b_s$~simulation is listed
  in Table~\ref{tab:tab_scaqua_1}. This table should be compared to Table~IV
  of Ref.~\cite{ajets}. From the fit parameters, we derive a cross section for 
  producing $b_s$ quarks that is ($11 \pm 2$)\% of the production cross
  section of $b$ quarks with the same mass, in reasonable agreement with the
  theoretical prediction mentioned in the previous section. 
  Table~\ref{tab:tab_scaqua_2} compares tagging rates in the data and in the
  simulation tuned according to the best fit. This table should be compared
  to Tables~VI and~IX of Ref.~\cite{ajets} and Table~III of 
  Ref.~\cite{sameside}.

  Because of the approximations
  in the modeling of the $b_s$ production and decay
  listed in Appendix~\ref{sec:ss-mod}, a determination of the 
  systematic errors of the $b_s$ production cross section is beyond the scope
  and reach of this study. As an example,
   the {\sc herwig} spectator model 
  predicts a reasonable, but also quite arbitrary, $\tilde{B}$ mass of 
  approximately $4\; \gevcc$.
  A $\pm 0.3 \; \gevcc$ change of the $\tilde{B}$ mass 
  results in a $\pm 40$\% change of the $b_s$ production cross section 
  returned by the best fit to the data.

\begin{figure}
 \begin{center}
 \leavevmode
 \includegraphics*[width=\textwidth]{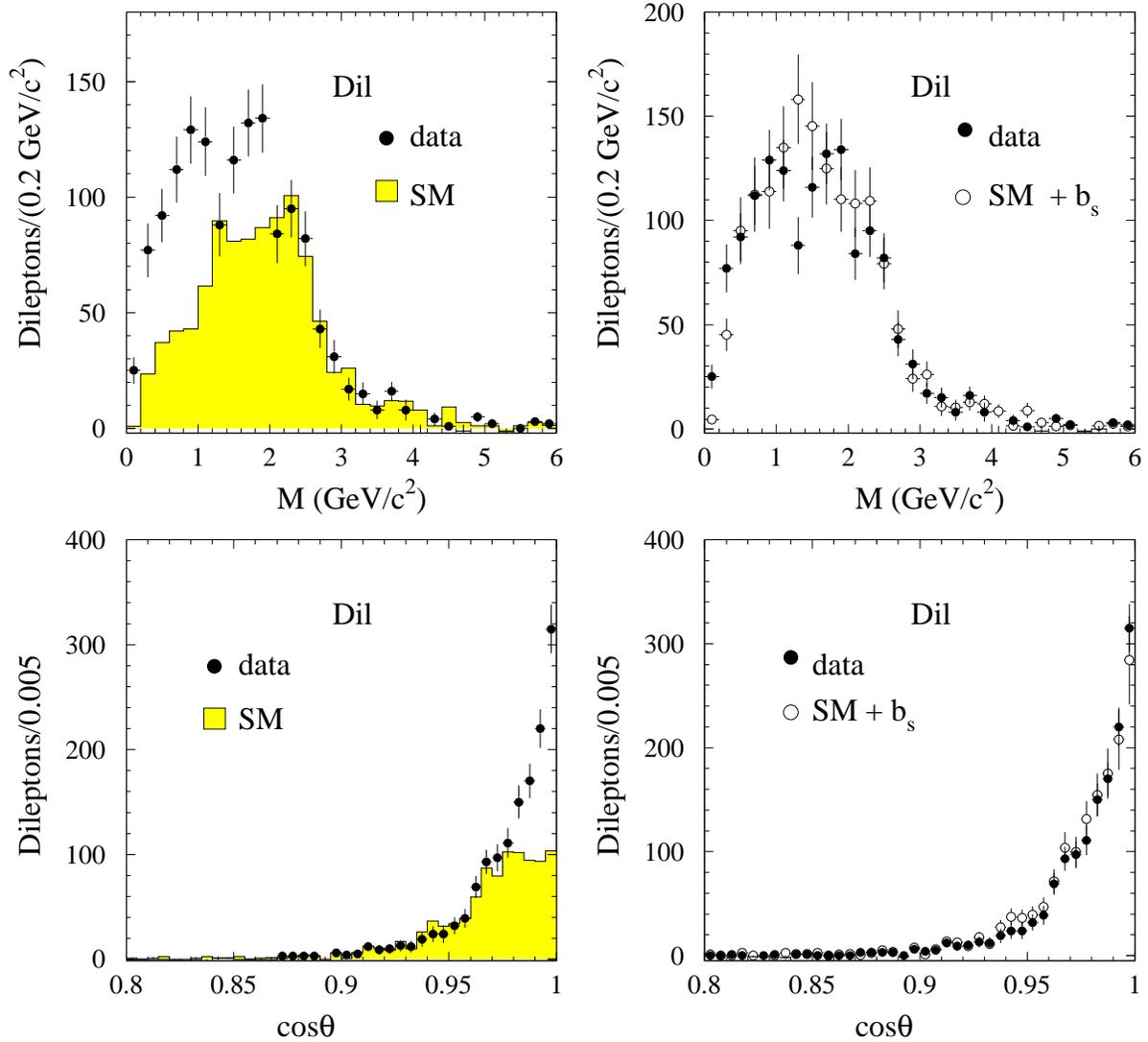}
 \caption[]{Observed distributions of the invariant mass, $M$, and the opening
            angle, $\theta$, of dileptons in the same jet are compared to 
            the simulation without (SM) and with $b_s$-quark production. 
            The simulation is normalized  with the fit parameters listed  
            in Table~\ref{tab:tab_scaqua_1}. }
 \label{fig:fig_scaqua_2}
 \end{center}
 \end{figure}

\begin{figure}
 \begin{center}
 \leavevmode
 \includegraphics*[width=\textwidth]{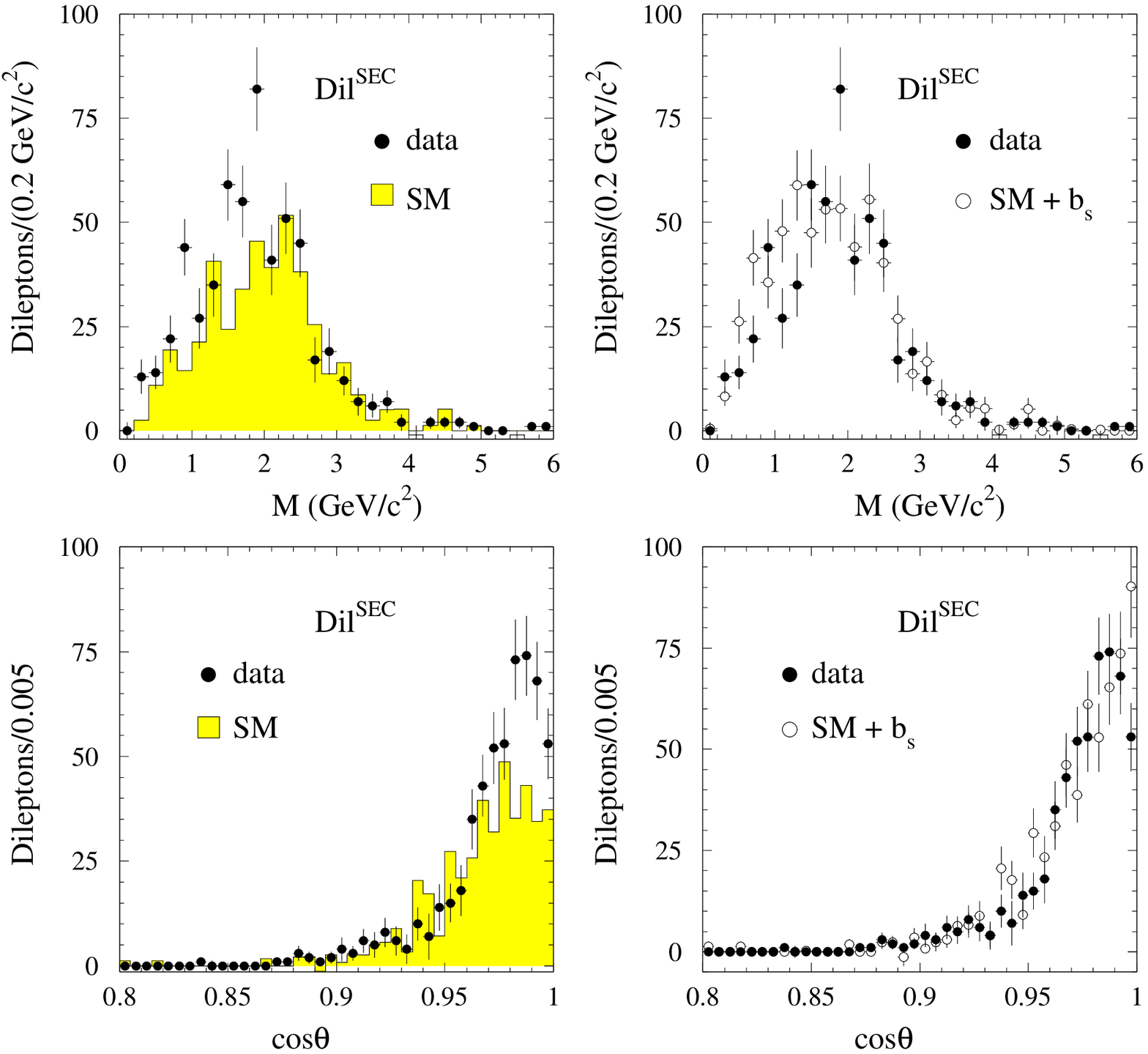}
 \caption[]{Observed distributions of the invariant mass $M$ and the opening
            angle $\theta$ of lepton pairs contained in a jet tagged by the 
            SECVTX algorithm are compared to the simulation.}
 \label{fig:fig_scaqua_3}
 \end{center}
 \end{figure}

\begin{figure}
 \begin{center}
 \leavevmode
 \includegraphics*[width=\textwidth]{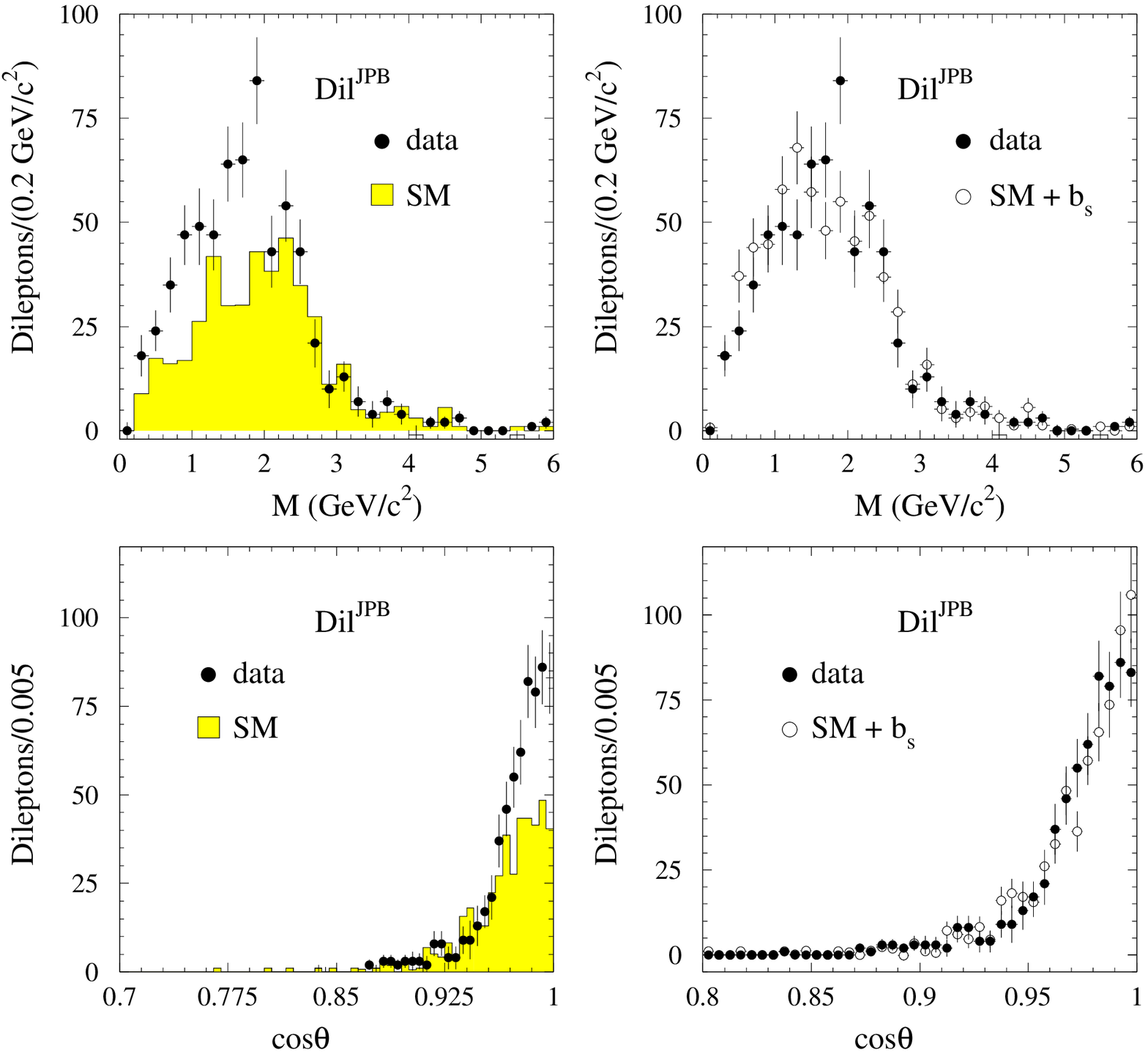}
 \caption[]{Observed distributions of the invariant mass $M$ and the opening
            angle $\theta$ of lepton pairs contained in a jet tagged by the 
            JPB algorithm are compared to the simulation.}
 \label{fig:fig_scaqua_4}
 \end{center}
 \end{figure}

\begin{figure}
 \begin{center}
 \leavevmode
 \includegraphics*[width=\textwidth]{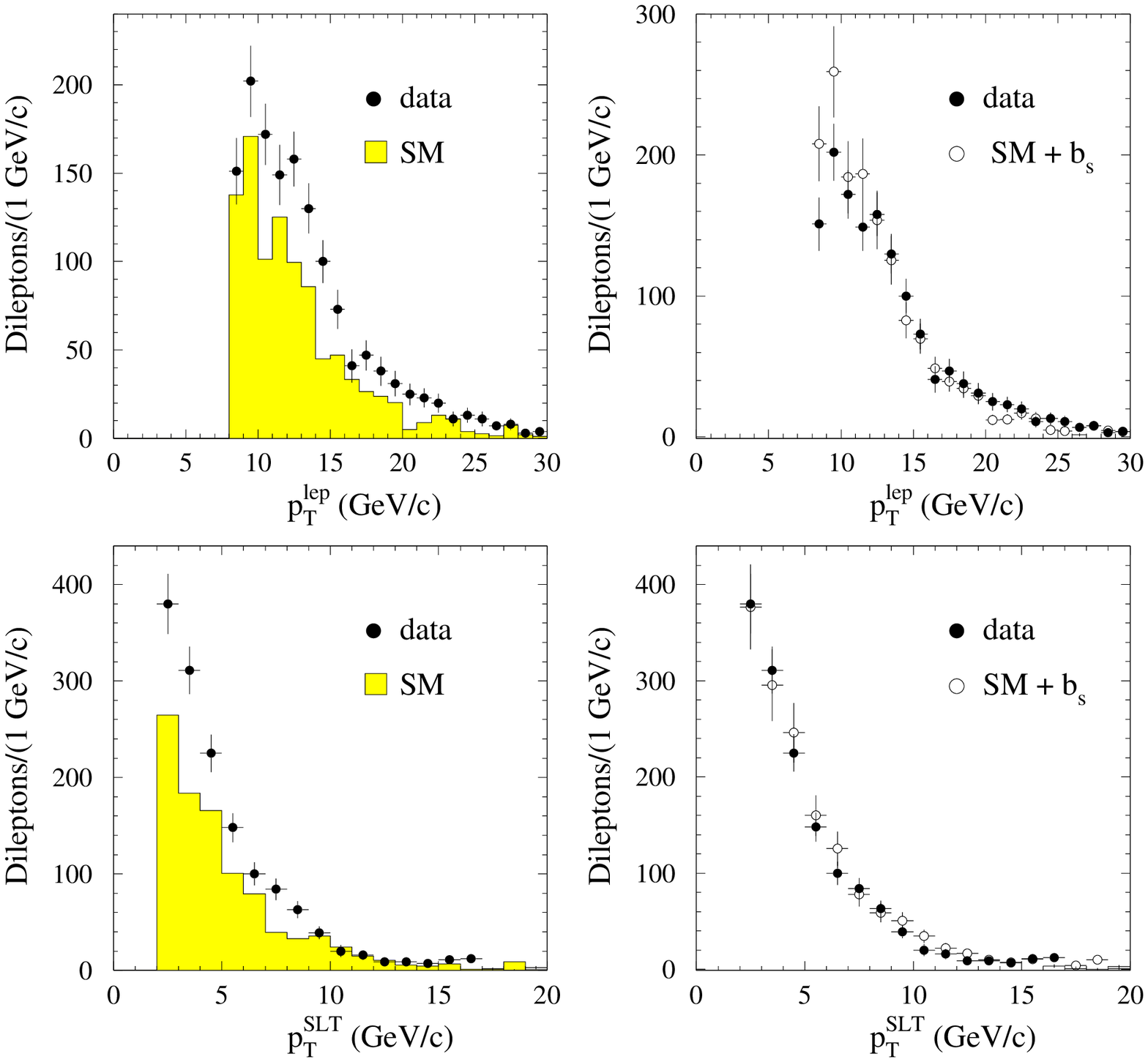}
 \caption[]{Transverse momentum distributions of the trigger (lep) and soft
            (SLT) leptons contained in the same jet are compared to the
             simulation.}
 \label{fig:fig_scaqua_5}
 \end{center}
 \end{figure}

  Figures~\ref{fig:fig_scaqua_2} to~\ref{fig:fig_scaqua_4} compare observed 
  distributions of the invariant mass and the opening angle of dileptons in
  the same jet to the  SM+$b_s$ simulation. The simulation is normalized with
  the fit parameters listed in Table~\ref{tab:tab_scaqua_1}. These data 
  appeared to be anomalous when compared to the SM simulation in 
  Ref.~\cite{sameside}. For completeness, Figure~\ref{fig:fig_scaqua_5}
  compares the transverse momentum distribution of dileptons in the same jet.
  In conclusion, the hypothesis of a light parton with a $100$\% semileptonic
  branching ratio and a production cross section typical of scalar quarks
  fits the yet unresolved discrepancies between the heavy flavor production
  observed at the Tevatron and the prediction of  conventional-QCD
  simulations reported in Refs.~\cite{ajets,derw,yale,d0-yale,sameside}.   
 \begin{table}
 \caption{Result of the fit with the SM simulation implemented with the pair 
          production of  scalar quarks $b_s$. The
          scalar quark mass is set to  $3.65\; \gevcc$ and its lifetime to 1.0
          ps. The fit yields a $\chi^{2}$ of 10 for 20 degrees of freedom.}
 \begin{center}
 \begin{ruledtabular}
 \begin{tabular}{lc}
 SECVTX scale factor ($b$ quarks)  &  $0.98 \pm 0.03$\\
 SECVTX scale factor ($c$ quarks)  &  $1.02 \pm 0.22$\\
 JPB scale factor    ($b$ and $c$) &  $1.00 \pm 0.02$\\
 $e$  norm.                        &  $0.96 \pm 0.06$\\
 $\mu$  norm.                      &  $0.96 \pm 0.06$\\
 $c$ dir. prod.                    &  $1.01 \pm 0.14$\\
 $b$ flav. exc.                    &  $0.79 \pm 0.16$\\
 $c$ flav. exc.                    &  $0.87 \pm 0.26$\\
 $g \rightarrow b\bar{b}$          &  $1.32 \pm 0.18$\\
 $g \rightarrow c\bar{c}$          &  $1.36 \pm 0.34$\\
 $b_s$ dir. prod.                  &  $0.07 \pm 0.03$\\
 $b_s$ flav. exc. + GSP            &  $0.20 \pm 0.04$\\
 \end{tabular}
 \end{ruledtabular}
 \end{center}
 \label{tab:tab_scaqua_1}
 \end{table}

 {\squeezetable
 \begin{table}
 \caption{Numbers of tags due to heavy flavors in the data and in the SM+$b_s$
          simulation. The simulation is normalized according to the fit result
          listed in Table~\ref{tab:tab_scaqua_1}. The SM ($b$ and $c$ quarks)
          and $b_s$ contributions are also listed separately. The electron 
          (muon) sample consists of events in which the trigger lepton is an
          electron (muon). In the data, a fraction of the trigger leptons is
          faked by hadrons that mimic the lepton signature: the heavy flavor
          purity determined by the fit is ($44.5 \pm 1.3$)\% for the electron
          sample and ($58.8 \pm 2.2$)\% for the muon sample.}
 \begin{center}
 \begin{ruledtabular}
 \begin{tabular}{lccccc}
  & & \multicolumn{3}{c}{\bf Electron sample} & \\
  Tag type      & Data & Simulation& SM& $b_s$-dir& $b_s$-(f.exc + GSP)\\
 $HF_{l-jet}$   
&$68544$
&$30478.0\pm 878.4$ &$26040.0\pm1282.2$ &$  575.7\pm 184.9$ &$ 3862.4\pm 882.7$\\
$HF_{a-jet}$
&$73335$
&$32559.0\pm 943.7$ &$27942.2\pm1354.9$ &$  593.1\pm 190.5$ &$ 4023.7\pm 919.5$\\
$HFT_{l-jet}^{SEC}$
&$10115.3 \pm   101.7$
&$10135.9\pm 144.7$ &$ 8641.1\pm 328.2$ &$  189.0\pm  61.4$ &$ 1305.8\pm 302.6$\\
$HFT_{l-jet}^{JPB}$
&$11165.4 \pm 115.8$
&$11136.1\pm 142.1$ &$ 9279.5\pm 379.7$ &$  235.0\pm  75.9$ &$ 1621.7\pm 372.1$\\
$HFT_{a-jet}^{SEC}$
&$3719.8 \pm 93.5$
&$ 3678.2\pm 102.8$ &$ 3384.1\pm 124.4$ &$  173.8\pm  56.5$ &$  120.3\pm  28.8$\\
$HFT_{a-jet}^{JPB}$
&$4021.3 \pm  140.7$
&$ 3972.5\pm 104.1$ &$ 3572.0\pm 130.4$ &$  228.2\pm  73.7$ &$  172.2\pm  40.4$\\
$HFDT^{SEC}$
&$1375.2 \pm   37.6$
&$ 1373.4\pm  55.9$ &$ 1279.0\pm  59.5$ &$   53.9\pm  18.0$ &$   40.5\pm  10.7$\\
$HFDT^{JPB}$
&$1627.8 \pm 43.7$
&$ 1647.2\pm  53.6$ &$ 1458.9\pm  57.8$ &$   89.7\pm  29.4$ &$   98.6\pm  23.7$\\
$HFT_{a-jet}^{SLT}$ 
&$862.2\pm   114.8$
&$  840.8\pm  98.1$ &$  534.7\pm  63.0$ &$  172.0\pm  58.0$ &$  134.1\pm  34.3$\\
$HFT_{a-jet}^{SLT \cdot SEC}$
&$322.1 \pm 23.3 $ 
&$  309.6\pm  26.9$ &$  219.8\pm  20.7$ &$   53.1\pm  17.7$ &$   36.6\pm   9.6$\\
$HFT_{a-jet}^{SLT \cdot JPB}$
&$349.4  \pm  26.3$ 
&$  353.0\pm  29.9$ &$  224.5\pm  20.1$ &$   69.2\pm  22.9$ &$   59.3\pm  14.8$\\
$Dil$
&$ 1111   \pm   54.2$
&$ 1149.6\pm 130.7$ &$  752.2\pm  88.0$ &$   52.2\pm  17.8$ &$  345.2\pm  86.9$\\
$Dil^{SEC}$
&$401.0 \pm   25.0$
&$  436.7\pm  35.9$ &$  312.5\pm  29.1$ &$   14.2\pm   4.9$ &$  109.9\pm  27.1$\\
$Dil^{JPB}$
& $   464.0 \pm   26.8$ 
&$  480.6\pm  38.9$ &$  323.1\pm  29.0$ &$   18.1\pm   6.1$ &$  139.4\pm  33.7$\\
 &  & \multicolumn{3}{c}{\bf Muon sample} &\\
 Tag type & Data & Simulation& SM& $b_s$-dir& $b_s$-(f.exc + GSP)\\
$HF_{l-jet}$
&$14966$
&$ 8796.6\pm 310.8$ &$ 7161.1\pm 425.7$ &$  234.2\pm  75.1$ &$ 1401.3\pm 311.1$\\
$HF_{a-jet}$
&$16460$
&$ 9557.2\pm 335.5$ &$ 7831.5\pm 456.7$ &$  246.5\pm  79.0$ &$ 1479.1\pm 328.4$\\
 $HFT_{l-jet}^{SEC}$
 &$3657.3\pm 60.8$
 &$3635.2\pm 81.2$  & $2962.3\pm 137.5$ & $ 94.9\pm 30.8$ &$ 578.0\pm 130.2$\\
 $HFT_{l-jet}^{JPB}$
 &$4068.6\pm 66.2$
 &$4051.1\pm 83.6$  & $3234.5\pm 156.8$ & $115.3\pm 37.2$ &$ 701.3\pm 156.3$\\
 $HFT_{a-jet}^{SEC}$
 & $943.8\pm 35.2$
 & $957.2\pm 38.7$  & $842.3\pm 43.5$   &$ 74.5\pm 24.2$& $  40.4\pm  9.7$\\
 $HFT_{a-jet}^{JPB}$
 &$1086.8\pm 45.0$
 &$1052.0\pm 40.0$  & $899.2\pm 45.2$   &$ 95.8\pm 30.9$& $  56.9\pm 13.2$\\
 $DT^{SEC}$
 &$ 452.6\pm 21.6$
 &$ 467.4\pm 27.6$  & $416.6\pm 28.5$   &$ 30.3\pm 10.1$& $  20.5\pm  5.3$\\
 $DT^{JPB}$
 &$ 546.4\pm 25.1$
 &$ 557.3\pm 25.5$  & $468.8\pm 26.9$   &$ 46.5\pm 15.2$& $  42.1\pm 10.0$\\
 $HFT_{a-jet}^{SLT}$ 
 &$ 271.9\pm 34.9$
 &$ 235.3\pm 31.3$  & $127.6\pm 18.2$   &$ 67.8\pm 22.9$& $  39.8\pm 10.2$\\
 $HFT_{a-jet}^{SLT \cdot SEC}$
 &$  63.2\pm 10.0$
 &$  82.3\pm 10.4$  & $ 46.8\pm 7.5$    &$ 23.8\pm 7.9$ & $  11.7\pm  3.2$\\
 $HFT_{a-jet}^{SLT \cdot JPB}$
 & $ 103.0\pm 12.4$
 & $ 101.9\pm 11.8$ & $ 54.9\pm 7.7$    &$ 29.9\pm 9.9$ & $  17.0\pm  4.3$\\
 $Dil$
 & $ 336.0\pm 28.5$  
 & $ 371.1\pm 45.9$ & $ 232.9\pm 31.6$  &$ 23.1\pm 7.9$ & $ 120.8\pm 29.9$\\
 $Dil^{SEC}$ 
 & $ 158.2\pm 15.9$
 & $ 168.3\pm 17.2$ & $ 116.9\pm 15.0$ &$ 8.0\pm 2.7$   & $  43.4\pm 10.6$\\
 $Dil^{JPB}$
 & $ 171.5\pm 16.5$
 & $ 177.7\pm 17.1$ & $115.5\pm 13.9$ &$ 9.5\pm 3.2$   & $  52.7\pm 12.6$\\
 \end{tabular}
 \end{ruledtabular}
 \end{center}
 \label{tab:tab_scaqua_2}
 \end{table}
}

 \section{Sensitivity of the data to other models} \label{sec:ss-inv}
  In the previous section we have used a $b_s$ mass of $3.65\; \gevcc$.
  This value was chosen because, in the invariant mass spectrum of muon 
  pairs collected by the CDF experiment, we have seen an excess of events
  at a mass of $7.22\; \gevcc$  consistent with the production of
  a bound state of scalar quarks. Since this excess could be a statistical 
  fluctuation, we have also generated several simulations with $b_s$ masses
  ranging from $2.5$ to $5.0\; \gevcc$. We find that the data are reasonably
  well 
  modeled when using any $b_s$ mass between 3 and $4\; \gevcc$.
 
  As correctly pointed out in Ref.~\cite{squa-3}, the measured semileptonic
  branching fraction of $b$ hadrons would be higher than the standard model
  expectation if $m_{b_s}+m_{\nu_s} \leq m_b$~\footnote{
  In this case the decay $b \rightarrow b_s \bar{\nu} \nu_s$ with
  $b_s \rightarrow l X$ is allowed and does not suffer CKM suppression.
  This decay would however produce leptons with  momentum significantly softer
  than those from direct semileptonic $b$ decays.}.
  We have produced simulations of the $b_s$ decay with $\nu_s$ masses of the
  order of 1 $\gevcc$. The predicted distributions for the dilepton kinematics
  look very similar to the ones shown in Figs.~\ref{fig:fig_scaqua_2}
  to~\ref{fig:fig_scaqua_4}; the fit to the observed tagging rates yields a
  higher $b_s$ production cross section (17\% of that for producing $b$ quarks 
  with the same mass).

  We have chosen a $b_s$ semileptonic decay mediated by the right-handed 
  component of the Higgsino coupling. We have also explored the use of a
  decay matrix mediated by the chargino left-handed coupling to ordinary
  matter (see Appendix~\ref{sec:ss-matr}) and the hypothesis of a $R$-parity
  violating $b_s \rightarrow l\; c$ decay. Both models produce leptons with a
  stiffer energy spectrum than the right-handed decay mode, but also provide
  a fair description of the data when using $b_s$ masses approximately equal
  or slightly smaller than 3.2 $\gevcc$. In these cases, the $b_s$ production 
  cross section returned by the best fit to the observed tagging rates is 
  $\simeq 5$\% of the production cross section of quarks with the same mass.
   We note that a
  sbottom quark with a mass of approximately 3.2 $\gevcc$ is 
  not favored by the
  study in Ref.~\cite{dimu}, but is out of the mass range
  investigated in the CLEO 
  analysis~\cite{cleo-sb}. In the next two subsections, we present two 
  additional tests of our model.
 \subsection{Study of  the {\boldmath $l^- \; D^0$} system}\label{sec:ss-ld0}
\begin{figure}
 \begin{center}
 \leavevmode
 \includegraphics{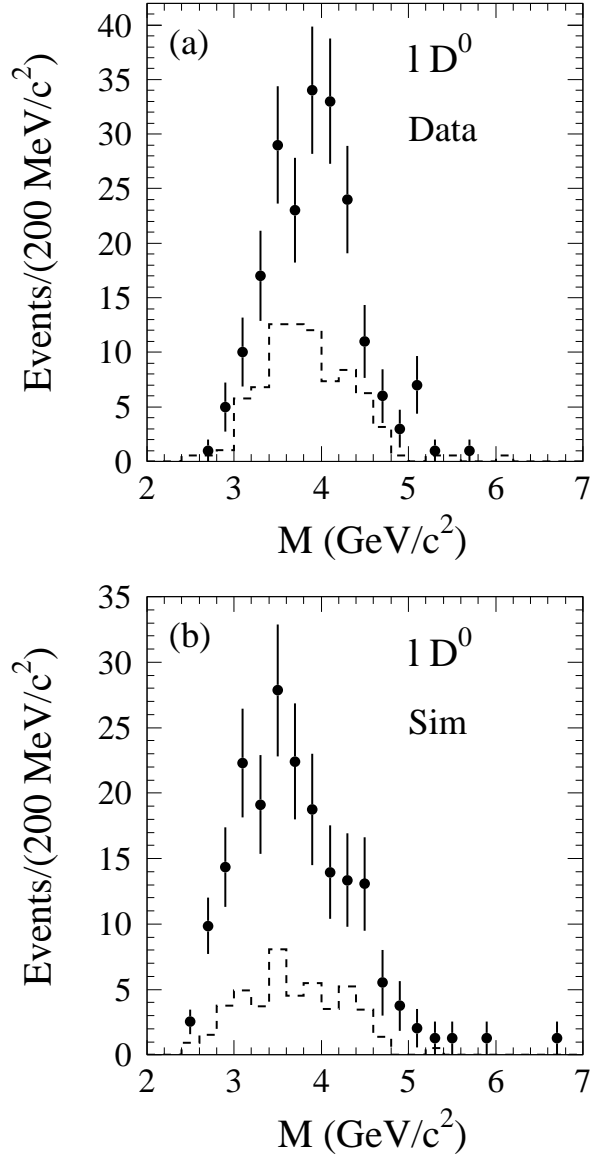}
 \caption[]{Invariant mass distributions of $l^- \;D^0$ candidates in (a) the
            data and (b) the SM+$b_s$ simulation. The dashed histograms 
            represent the background evaluated using wrong-sign combinations
            (see text).}
 \label{fig:fig_ld0_2}
 \end{center}
 \end{figure}

\begin{figure}
 \begin{center}
 \leavevmode
 \includegraphics{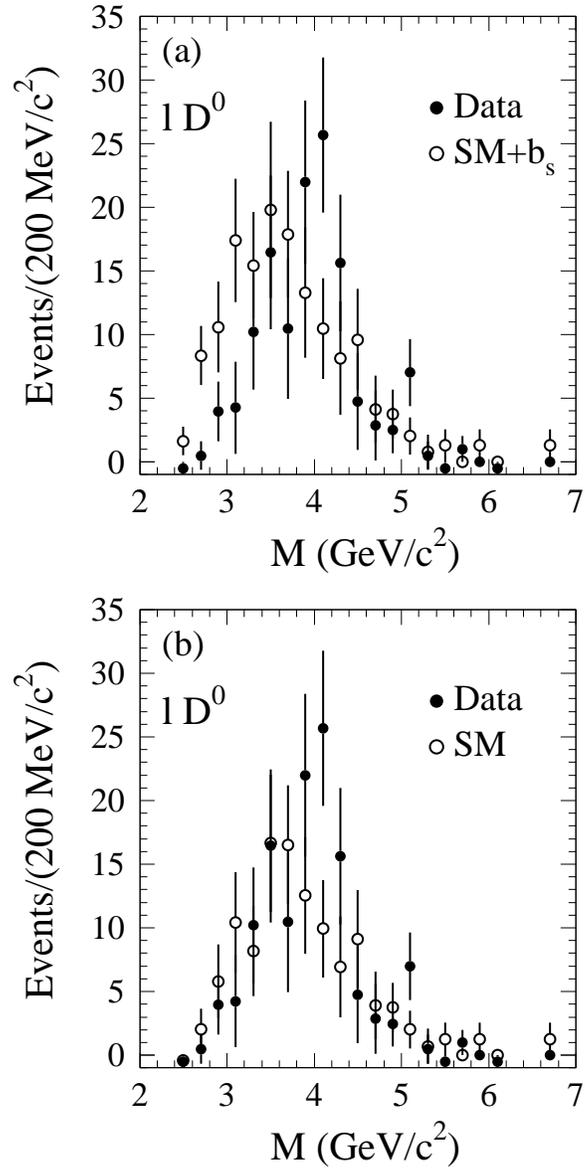}
 \caption[]{The distribution of the $l^- \; D^0$ invariant mass after 
            background removal is compared to the prediction of (a) the 
            SM+$b_s$ simulation and (b) the SM contribution only.}
 \label{fig:fig_ld0_3}
 \end{center}
 \end{figure}

  In Sec.~X~B of Ref.~\cite{ajets}, the $b$ purity of the inclusive lepton
  sample has been verified by counting the number of $D^0$ mesons accompanying
  the trigger leptons. In the simulation, most of the $l^- \; D^0$ (and its
  charge conjugate) pairs  arise from decays of single $b$ hadrons. In that
  study, the ratio
  of the number of identified  $l^- \;  D^0$ pairs in the data to that in the
  SM simulation is $0.92 \pm 0.13$. This agreement has favored the initial 
  assumption that $b_s$ quarks decay into $c$ quarks because
  our model replaces a fraction
  of the $b\bar{b}$ production with $b_s \bar{b}_s$ production. We search the 
  SM+$b_s$ simulation for $D^0 \rightarrow K^- \pi^+$ candidate around the 
  direction of the trigger lepton
  following the
  same procedure described in Sec.~X~B of Ref.~\cite{ajets}.
 In the mass range of $1.82-1.92\; \gevcc$
  we identify $147 \pm 15$ $ D^0$ mesons (after subtracting a background of 
  $47 \pm 6$ events), in agreement with the number $126 \pm 15$ found in the
  data (after removing a background of $79 \pm 6$ events). In the SM+$b_s$ 
  simulation $24$\% of the $l^- \;D^0$ pairs arise from $b_s$ decays. Since
  the $b_s$ quark is assumed to be lighter than a $b$ quark, $l^- \;  D^0$ 
  pairs from $b_s$ decays should cluster at smaller invariant masses than 
  those from $b$ decays. It is therefore of interest to compare the 
  $l^- \;D^0$ invariant mass distribution in the data and the simulation
  (see Fig.~\ref{fig:fig_ld0_2}). In this comparison the shape of the
  background is evaluated by using wrong-sign combinations ($l^- K^+ \pi^-$)
  with the $K^+ \pi^-$ invariant mass in the range $1.82-1.92\; \gevcc$. 
  Figure~\ref{fig:fig_ld0_3}(a) compares $l^- \; D^{0}$ invariant mass
  distributions in the data and in the simulation
  after background removal~\footnote{
   Because of the $p_T$ cuts used to select
 leptons, pions, and kaons, the $p_T$ threshold of the $l \; D^0$ system
  decreases rapidly with increasing  $l \; D^0$ invariant mass. Therefore, the
  $l^- \; D^0$ invariant mass distribution is also affected by the $p_T$ distribution
  of the $\tilde{B}$ mesons that in turn is affected by the shortcomings
 of the {\sc herwig}
  fragmentation and hadronization model (see Appendix~\ref{sec:ss-mod}).}.

  A comparison of the two distributions in the mass range $2.4-5.2\; \gevcc$
  yields a $\chi^2$ of 10 for 13 DOF. However, the SM part of the simulation
  contains $112 \pm 14$ $l^- \;D^0$ candidates, which is also in agreement
  with the data ($126 \pm 15$ candidates). The invariant mass distribution of
  the $l^- \;D^0$ candidates contributed by the SM part of the simulation,
  shown in Figure~\ref{fig:fig_ld0_3}(b), also agrees  with the data
  (the $\chi^2$ of the comparison of the invariant mass distributions is
  6 for 13 DOF). In the limited statistics of this sample, the
   $l^- \; D^0$  data  do not provide additional support to our model.

 \newpage 
 \subsection{Rates of lepton-jets containing $J/\psi$ mesons} 
 \label{sec:ss-psi}
  $J/\psi$ mesons are produced in the decay of $B$ hadrons but not in the
  decay  of
  $\tilde{B}$ hadrons with a $100$\% semileptonic branching ratio. The
   SM+$b_s$ simulation
  predicts a rate of $J/\psi$ mesons from $B$
  decays that is approximately 20\% smaller than that predicted by the SM simulation
  (see Table~XIV of Ref.~\cite{ajets}).
   Rates of  $J/\psi$ in the data and in the  SM+$b_s$ simulation
   are listed in Table~\ref{tab:tab_psi_1}.
   Rates of $J/\psi$ mesons contained in lepton-jets with
  SECVTX and JPB tags arise only from $B$ decays, and are well predicted by
  the SM+$b_s$ simulation. Before tagging, ($50\pm 5$)\% 
  of the $J/\psi$ mesons identified in the data
   (first row of Table~\ref{tab:tab_psi_1}) 
  do not arise from  $B$ decays~\cite{ajets,bprompt}. 
    The SM+$b_s$ prediction  (180$\pm$20) is still slightly larger
    than the data in which $130\pm 11$ 
     $J/\psi$ mesons originate from $B$ decays.
  \begin{table}
 \caption{Numbers of $J/\psi$ mesons in the data and in the SM+$b_s$ 
          simulation.}
 \begin{center}
 \begin{ruledtabular}
 \begin{tabular}{lcccc}
 &\multicolumn{2}{c}{ $J/\psi \rightarrow e^+ e^-$ }&
\multicolumn{2}{c}{  $J/\psi \rightarrow \mu^+ \mu^-$}\\
                    & data              &  simulation     & data &simulation\\
 $Dil_{\psi}$       & $ 176.0 \pm 14.4$ & $ 131.4\pm 18.7$ & $83.0\pm 9.4$
                    & $  48.3 \pm 8.6 $ \\
 $Dil^{SEC}_{\psi}$ & $  57.8 \pm 8.8 $ & $  61.3\pm 9.5 $ & $31.9\pm 5.8$  
                    & $  25.7 \pm 5.2 $ \\
 $Dil^{JPB}_{\psi}$ & $  61.2 \pm 8.4 $ & $  57.5\pm 8.3 $ & $29.6\pm 5.7$ 
                    & $  28.1 \pm 5.2 $  \\
 \end{tabular}
 \end{ruledtabular}
 \end{center}
 \label{tab:tab_psi_1}
 \end{table}

 \section{Lifetime}~\label{sec:ss-scalifet}
\begin{figure}[htb]
 \begin{center}
 \leavevmode
 \includegraphics*[width=\textwidth]{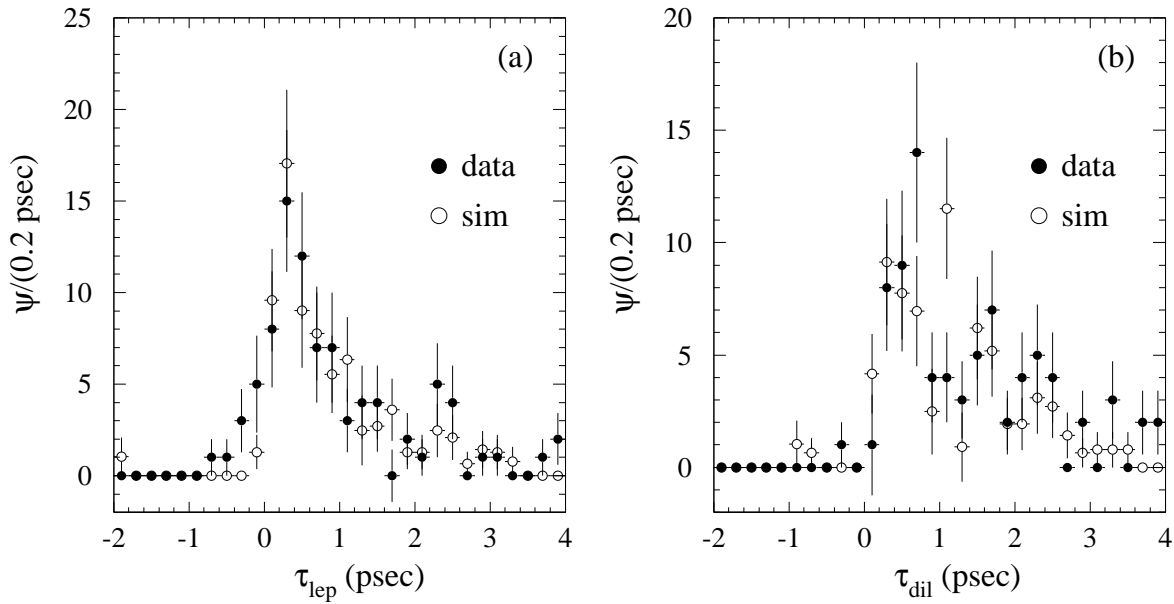}
 \caption[]{Test of the distributions used to estimate the $b_s$ lifetime using
            $J/\psi$ mesons contained in a jet tagged by the JPB algorithm
            (see text).}
 \label{fig:psi_1}
 \end{center}
 \end{figure}

  In this section we use several techniques to estimate the lifetime of the
  object producing the excess of soft lepton tags with respect to the SM 
  simulation. We extract the lifetime by a data to simulation comparison
  of the distributions of the following quantities strictly related
  to the lifetime:
 \begin{enumerate}
 \item $\tau_{\rm dil}= \frac{L_{xy} \; M} {c \; p_T}$,
       where $M$ and $p_T$ are the invariant mass and transverse momentum of
       a lepton pair (trigger lepton plus soft lepton), and $L_{xy}$ is the
       projection on the $p_T$ direction of the distance between the primary
       vertex of the event and the vertex determined by the lepton tracks.
 \item $\tau_{\rm l}= \frac{4 \; \vec{d} \cdot \vec{p}_T}{\pi \; c \; |p_T| }$,
       where $\vec{d}$ is the vector corresponding to the distance of closest
       approach of a lepton track to the primary event vertex, and $\vec{p}_T$
       is the transverse momentum of the jet containing a trigger (lep) or 
       soft (SLT) lepton.
 \end{enumerate}
  We test these methods using lepton pairs consistent with $J/\psi$ mesons
  produced by $b$ decays ($Dil_{\psi}^{JPB}$ in Table~\ref{tab:tab_psi_1}). 
  Figure~\ref{fig:psi_1} shows agreement between data and simulation within
  the limited statistics of the $J/\psi$ sample.

\begin{figure}
 \begin{center}
 \leavevmode
 \includegraphics*[width=\textwidth]{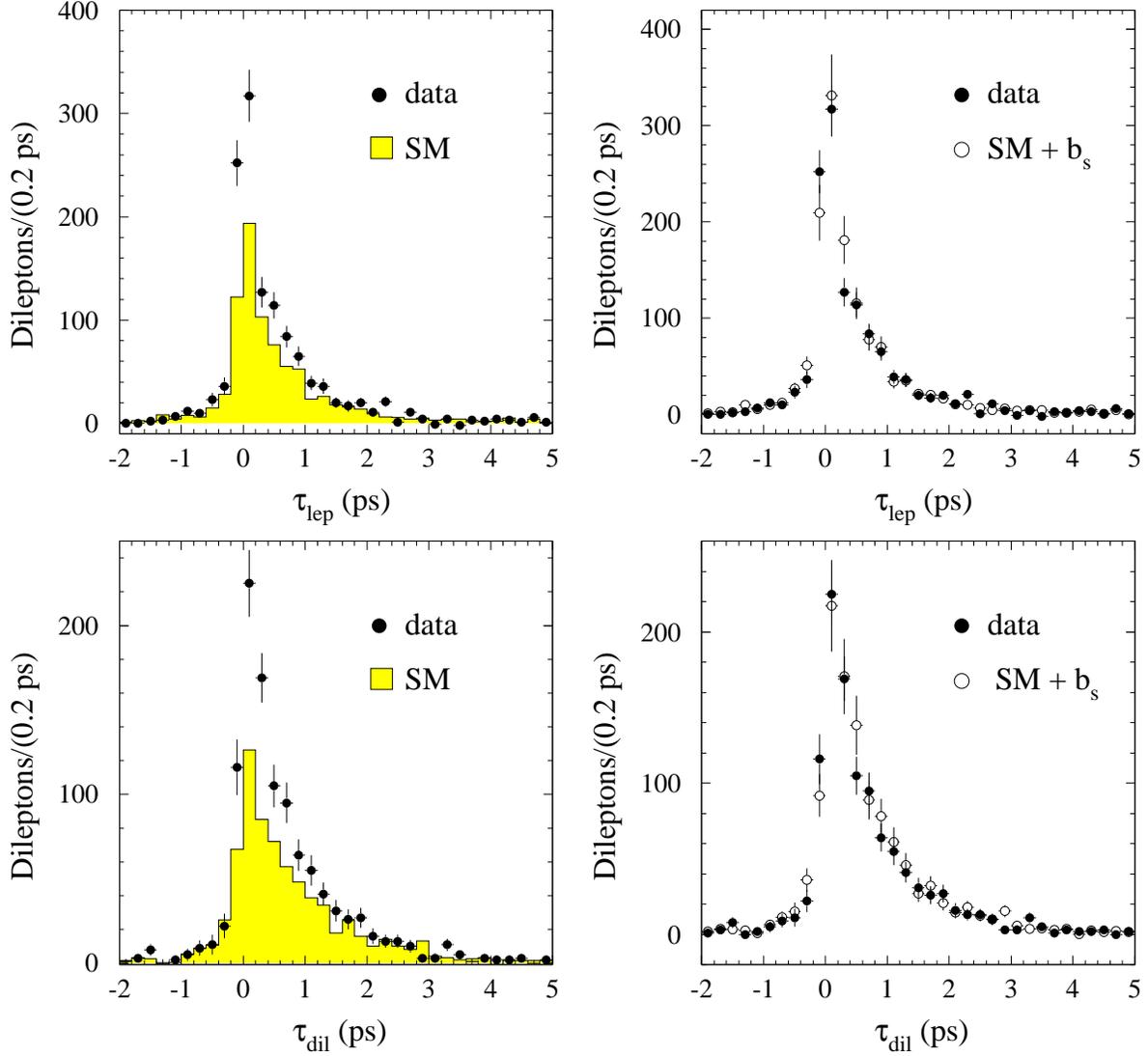}
 \caption[]{Lifetime distributions derived using lepton pairs in the same jet
            are compared to the prediction of the simulations without (SM) and
            with $b_s$ production. The simulation is normalized according to 
            the fit result listed in Table~\ref{tab:tab_scaqua_1}. A comparison
            of the measured and predicted $\tau_{\rm lep}$ distributions yields
            a $\chi^2$ of 8 for 15 DOF ($-1.0 \leq \tau_{\rm lep}\leq 2$ ps);
            the comparison of the $\tau_{\rm dil}$ distributions yields a
            $\chi^2 $ of 8 for 12 DOF ($-0.4 \leq \tau_{\rm dil}\leq 2$ ps).}
 \label{fig:fig_scalifet_1}
 \end{center}
 \end{figure}

\begin{figure}
 \begin{center}
 \leavevmode
 \includegraphics*[width=\textwidth]{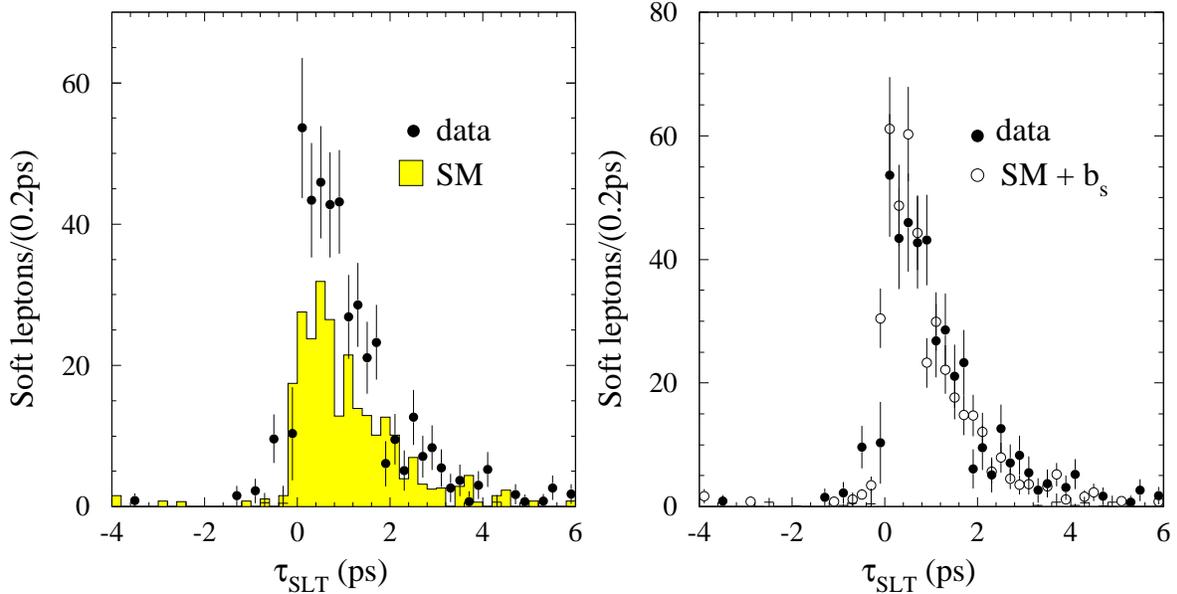}
 \caption[]{Lifetime distributions derived using soft leptons contained in
            away-jets with JPB tags. The comparison of the $\tau_{\rm SLT}$ 
            distributions in the data and the SM+$b_s$ simulation yields a
            $\chi^2 $ of 15 for 10 DOF ($0 \leq \tau_{\rm SLT} \leq 2$ ps).}
 \label{fig:fig_scalifet_2} 
 \end{center}
 \end{figure}

\begin{figure}
 \begin{center}
 \leavevmode
 \includegraphics{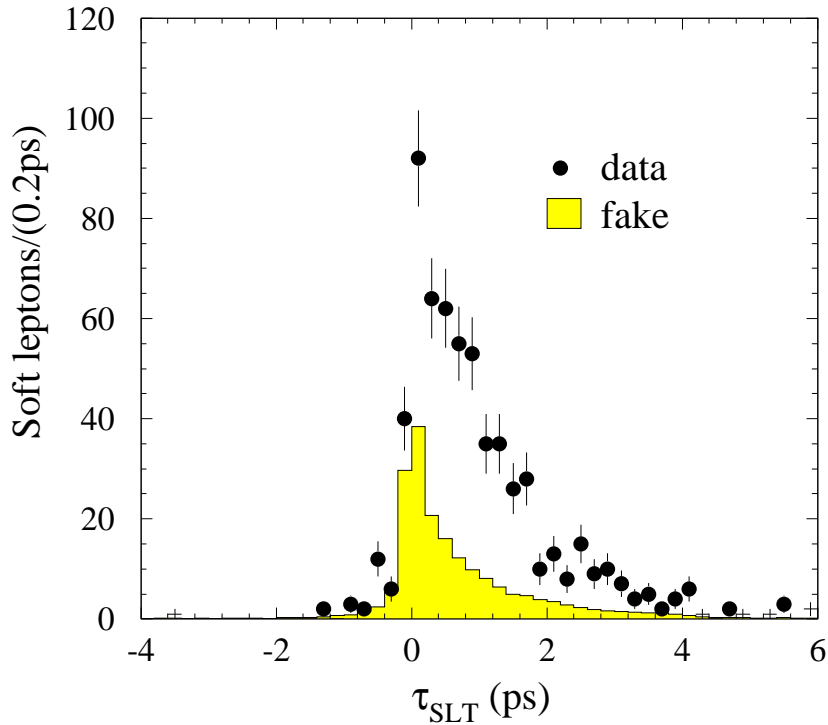}
 \caption[]{The lifetime distribution of soft leptons in away-jets tagged by
            JPB is compared to the expected contribution of fake SLT tags.}
 \label{fig:fig_scalifet_3}
 \end{center}
 \end{figure}

\begin{figure}
 \begin{center}
 \leavevmode
 \includegraphics*[width=\textwidth]{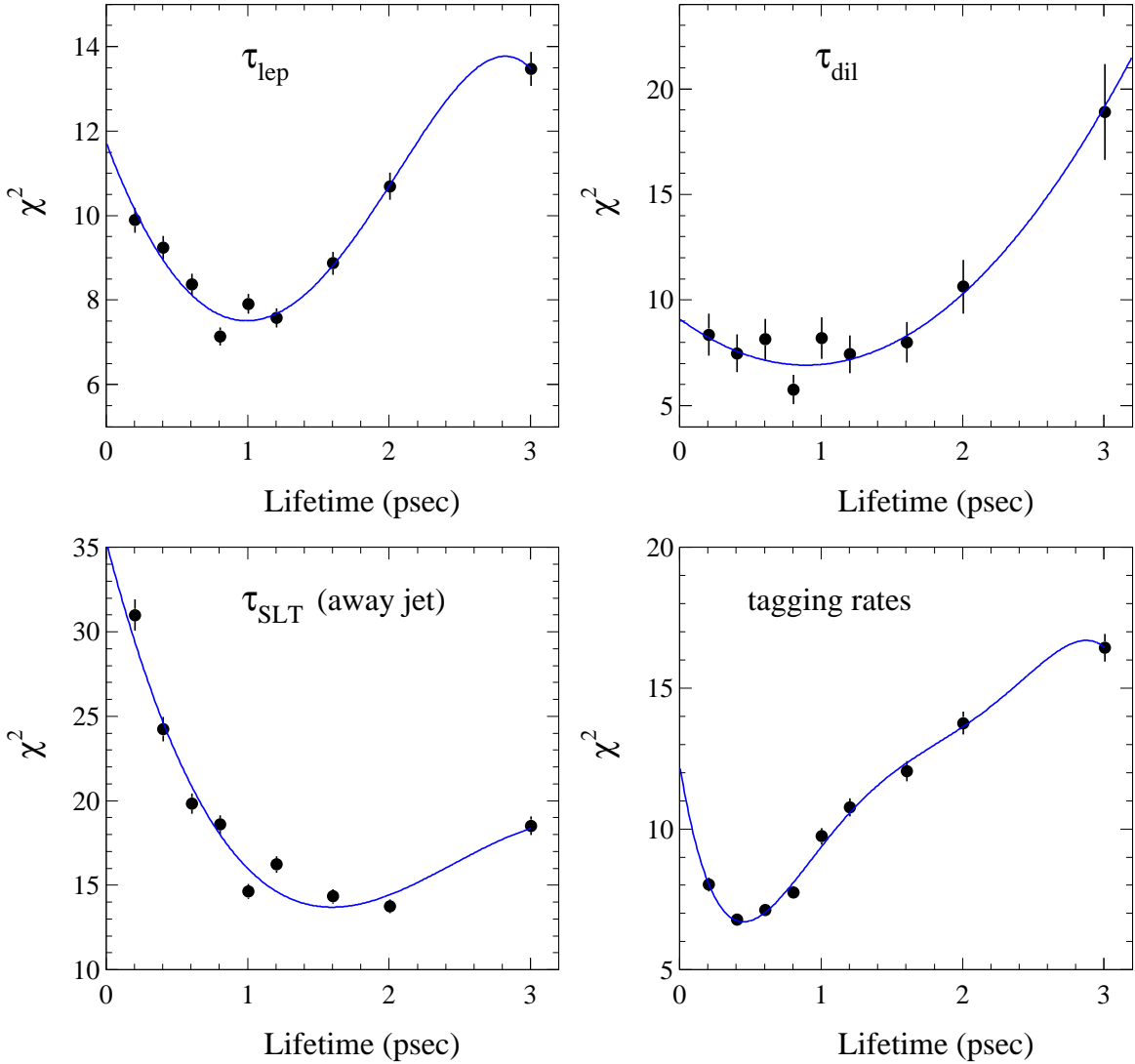}
 \caption[]{Yields of $\chi^{2}$ resulting from comparisons of data and
            simulation as a function of the $b_s$ lifetime.}
 \label{fig:fig_scalifet_4}
 \end{center}
 \end{figure}

  Figures~\ref{fig:fig_scalifet_1} and~\ref{fig:fig_scalifet_2} compare 
  lifetime distributions in the data and in the simulation implemented with
  the production of a scalar quark with a 1.0 ps lifetime and normalized as
  in Table~\ref{tab:tab_scaqua_1}. Using the track parameters of the trigger
  and soft leptons in the same jet we construct the quantities 
  $\tau_{\rm dil}$ and $\tau_{\rm lep}$. Using soft leptons in away-jets
  with JPB tags, we construct the quantity $\tau_{\rm SLT}$. In the data this
  last distribution has been obtained after subtracting the expected amount
  of fake SLT tags. The $\tau_{\rm SLT}$ distribution for fake tags is
  approximated using the distribution of all SLT candidate tracks in these
  jets and is shown in Fig.~\ref{fig:fig_scalifet_3}.
 For each distribution we evaluate the quantity
    \[  \chi^{2}= \sum_i \frac { (D(i) - S(i))^{2} }{ ED(i)^2+ES(i)^2} \]
 where $D(i)$ and $S(i)$ are the contents of the $i^{\rm th}$-bin of the 
 measured and predicted distributions, respectively ($ED(i)$ and $ES(i)$ are
 the corresponding errors). The $\chi^{2}$ is constructed avoiding large
 negative and positive tails which are not well modeled by our detector 
 simulation~\cite{cdf-tsig}. We have produced a set of scalar quark
 simulations with lifetime values ranging from $0.2$ to $3.0$~ps which we
 add to the SM~simulation. We first fit each of these
 simulations to the data following the same procedure used in Sec.~\ref{sec:ss-scaqua}. 
 Since some tagging rates depend upon the lifetime, we use the $\chi^{2}$
 value returned by the fit as an additional discriminant. The $\chi^{2}$ 
 yields as a function of the $b_s$ lifetime are shown in 
 Fig.~\ref{fig:fig_scalifet_4}. 
 
 The fit of the measured tagging rate yields a lifetime of $0.54 \pm 0.3$ ps

 The $\tau_{\rm lep}$ distribution yields a lifetime of $1.0 \pm 0.5$ ps

 The $\tau_{\rm dil}$ distribution yields a lifetime of $0.9 \pm 0.6$ ps

 The $\tau_{\rm SLT}$ distribution yields a lifetime of $1.55^{+0.6}_{-0.4}$ ps

 The average of the four results is 1.0 ps; the uncertainty defined as the RMS
 deviation of the four measurements is 0.4 ps.
 \section{Conclusions} \label{sec:concl}
  We review a number of known discrepancies~\cite{ajets,derw,yale,d0-yale,sameside}
  between the heavy-flavor properties of jets produced at the Tevatron
  and the prediction of conventional-QCD simulations. 
   These discrepancies may well be the result of  larger than
   estimated experimental difficulties hidden in all $b\bar{b}$ cross
   section measurements~\cite{bbar}. However, in this study, we
  entertain the possibility that these effects are real,
  and  investigate the constraints imposed by the data on 
  the hypothesis that a fraction of the 
  heavy flavor hadrons produced at the Tevatron might be due to
  unconventional sources.
  We postulate the existence of a sbottom quark $b_s$
  that decays to a lepton, a charmed quark, and a neutral scalar 
  non-interacting particle $\nu_s$ with a 100\% semileptonic branching 
  fraction, and implement its production into a conventional-QCD simulation 
  based on the {\sc herwig} Monte Carlo generator program. We initially use a
  $b_s$ mass of 3.6 $\gevcc$ since it is not excluded in the study of 
  Ref.~\cite{dimu}, a massless $\nu_s$, a lifetime of 1 ps, and a decay 
  mediated by the higgsino coupling to the right-handed matter. We use this
  generator to simulate the inclusive lepton sample collected by the CDF
  experiment in the $1992-1995$ collider run. The data sample consists of
  events with two or more jets, one of which is required to contain a lepton
  with $p_T \geq 8 \; \gevc$; this jet is referred to as lepton-jet, whereas
  jets recoiling against it are called away-jets. Following Ref.~\cite{ajets},
  this study uses rates of lepton- and away-jets containing displaced
  secondary vertices (SECVTX tags), tracks with large impact parameters
  (JPB tags), and  soft ($p_T \geq 2 \; \gevc$) leptons (SLT tags) in order
  to determine the bottom, charmed, and $b_s$ content of the data. 
  We tune the parton-level cross section predicted by the simulation within
  the experimental and theoretical uncertainties. The best fit to the data
  returns a $b_s$ production cross section at the Tevatron that agrees with
  the NLO prediction for pair production of scalar quarks.
  According to the 
  fit, the $b_s$ contribution to the inclusive lepton sample is approximately
  23\% (35\%) of that of the $b$ quark in the electron 
  (muon) sample. This contribution  models the invariant mass
  and opening angle distribution of lepton pairs contained in the same jet
  that were found anomalous when compared to a SM simulation
  in Ref.~\cite{sameside}. 
  We have studied the sensitivity of the data to slightly different models
  and investigated  the lifetime of the hypothetical $b_s$ quark.
  The best fit to the data replaces approximately 20\% of the $b$ production 
  with $b_s$
  production. In our hypothetical model, the invariant mass
  distribution of  $l \; D^0$ 
  pairs coming from $b_s$ decays is expected to peak at  smaller values than
  that of $l \; D^0$ 
  pairs coming from $b$ decays. In the limited statistics of the data, 
  we see no clear evidence of this. 
  Since  $b_s$ decays do not produce $J/\psi$ mesons,
  we have verified that rates of $J/\psi$ mesons in the data are consistent
  with those predicted by the simulation that includes $b_s$ production.
 \section{Acknowledgments}
   We thank the Fermilab staff, the CDF collaboration, and the technical staff
   of its participating Institutions for their contributions.
 We thank M.~Spira for making available the code of
 the {\sc prospino} Monte Carlo generator, and  M.~Mangano for working out many details
 of the matrix element used to decay sbottom quarks.
 This work was
   supported by the U.S.~Department of Energy and the Istituto Nazionale di
   Fisica Nucleare.
 \appendix
 \newcommand{\sbtolcn}{{\rm\bf b_s \rightarrow l c\nu_s}}
 \section{Decay $\sbtolcn$ } \label{sec:ss-matr}
  The decay of $b_s$ hadrons is modeled using the spectator model (routine
  {\sc hwdhvy}) available in the version 5.6 of {\sc herwig}. In this model, 
  {\sc herwig} normally weights the phase space of the three-body 
  semileptonic decay of a $b$ quark with the V-A matrix element. In the case
  of a scalar quark we have modified the {\sc herwig} code and weight the
  phase-space with the matrix element:
 \begin{eqnarray}
  \frac{d\Gamma}{dz_c dz_l}= K
                     [(1-z_c)(1-z_l)-R_{\nu_s}+R_c(z_c-z_l+R_{\nu_s}-R_c)]
 \end{eqnarray}
  where $K$ is a normalization factor, $R_c=m_c^2 / m_{b_s}^2$, and
   $R_{\nu_s}=m_{\nu_s}^2 / m_{b_s}^2$.

  Here $z_c$ and $z_l$ are defined as
   \[ z_c = \frac{2p_{b_s} \cdot p_c}{m_{b_s}^2}
    \mbox{~~ and~~~ }
    z_l = \frac{2p_{b_s} \cdot p_l}{m_{b_s}^2} \]
   The phase space limits are:
        \[ 2 \sqrt{R_c} < z_c < 1 + R_c - R_{\nu_s} \]
        \[\frac{1+R_c-R_{\nu_s}-z_c}{1-[z_c-\sqrt{z_c^2-4R_c}]/2} < z_l <
        \frac{1+R_c-R_{\nu_s}-z_c}{1-[z_c+\sqrt{z_c^2-4R_c}]/2} \]
  This matrix element for the decay $b_s \rightarrow c l \nu_s$ is derived
  from the tree level calculation outlined in Ref.~\cite{sq-koba}:
 \scriptsize
 \begin{eqnarray}
  {\cal M}= \sum_{j=1}^{2} \frac {-g^{2} V_{j1} F_{\rm L} |V_{c\tilde{b}}| }
  {  (p_{\tilde{b}}-p_c)^{2}-
  m_{\chi_{j}}^{2} }  \left[ \left[  U_{j1} cos \theta - \frac {m_{b}
  U_{j2}
  sin \theta}{\sqrt{2} m_{W} \cos \beta} \right] m_{\chi_{j}} \bar{u} (p_c)
  P_R v(p_l) +\frac {m_c V_{j2}^{*} cos \theta} {\sqrt{2} m_{W} \sin \beta}
  \bar{u}(p_c) {\not\!p_{\tilde{\nu}} } P_R v(p_l)   \right]
 \end{eqnarray}
 \normalsize
  where $\chi_{j}$'s  are the chargino mass eigenstates, and $U$ and $V^{*}$
  are the mixing matrices for the right and left-handed charginos, 
  respectively. The first subscript of $U$, $V$ corresponds to mass 
  eigenstates and the second to weak eigenstates (1 for the gaugino and 2 for
  the higgsino). Here tan$\beta$ is the ratio of the vacuum expectation 
  values of the two Higgs fields, $\theta$ is the mixing angle between
  left-handed and right-handed scalar quarks, $F_{\rm L}$ is the fraction of
  left-handed component of the scalar neutrino, and $|V_{c\tilde{b}}|$ is the
  CKM matrix element. Equation (A1) follows from (A2) when the decay is
  mediated by the higgsino coupling to the right-handed matter. If the decay
  is mediated by the gaugino coupling to the left-handed matter the matrix
  element is
  \begin{eqnarray}
    \frac {d\Gamma}{dz_c dz_l}= K (z_c+z_l -1+R_{\nu_s}-R_c ).
  \end{eqnarray}
  In the latter case the two fermions in the final state are both left-handed
  and tend to be produced back-to-back since the initial state is spinless.
  Using equation (A1) the two fermions in the final state have opposite
  handedness and tend to be produced more collinear than when using equation
  (A3). It follows that the matrix element (A1) produces leptons with a
  momentum distribution appreciably softer than that produced by the matrix 
  element (A3) or by a phase space decay; this latter mode has been
  used in Ref.~\cite{cleo-sb} to search evidence for such a scalar quark.
  
\begin{figure}[htb]
 \begin{center}
 \leavevmode
 \includegraphics*[width=\textwidth]{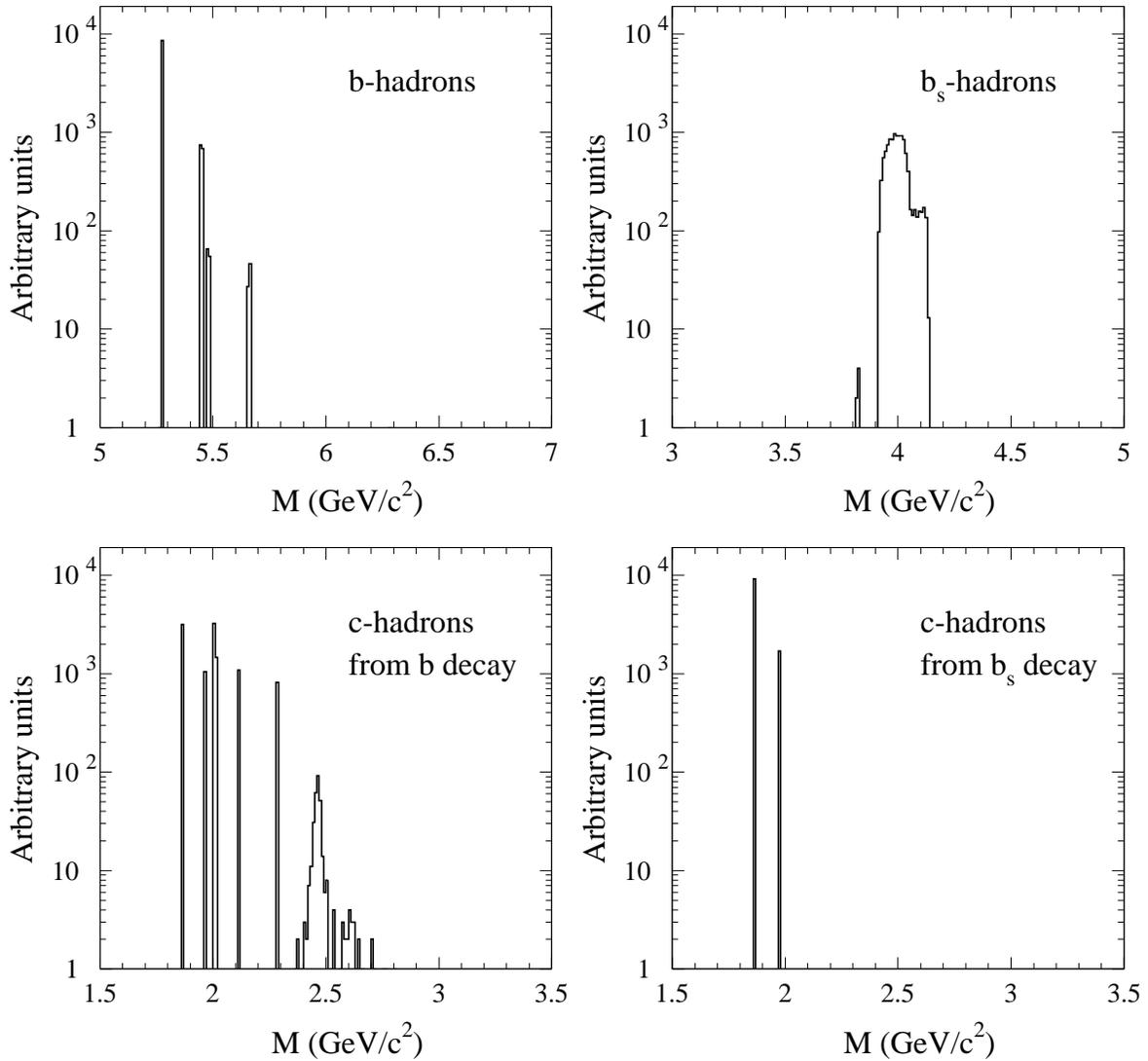}
 \caption[]{Mass spectrum of:  $b$ and $b_s$ hadrons formed by 
            the {\sc herwig} generator using  $m_b =4.75\; \gevcc$ 
            and  $m_{b_s} = 3.65\; \gevcc$; 
             $c$ hadrons from $b$-hadron decays
            modeled with the {\sc qq} generator; and 
             $c$ hadrons from $b_s$-hadron
            decays modeled with the {\sc herwig} spectator model.}
 \label{fig:fig_10.1}
 \end{center}
 \end{figure}

  The $c$ quark emerging from the $b_s$ decay is recombined with the 
  spectator quark by the {\sc herwig} spectator model. The decay of the
  resulting $c$ hadron is performed with the {\sc qq} Monte Carlo program.
  In the spectator model, excited $D$-meson states are produced only by the
  hadronic current carried by the virtual $W$ (so called {\em upper vertex}).
  Since we impose that the gauge fermion involved in the scalar quark decay
  has only leptonic decay modes, the simulation produces a very simple list
  of $c$ hadrons with respect to the {\sc qq} generator (see 
  Fig.~\ref{fig:fig_10.1}). The value of the $\tilde{B}$ masses returned by the
  spectator model, which does not have a look-up table similar to that for 
  $B$ hadrons, are no more than an educated guess, probably with an
  uncertainty of a few hundred MeV. The fractions of different $c$-hadrons
  produced by the spectator model are also quite arbitrary.
\newpage
 \newcommand{\sbot}{{\rm\bf b_s}}
 \section{Modeling of the $\sbot$ production} \label{sec:ss-mod}
  We implement the production of pairs of scalar quarks in $p\bar{p}$ 
  collisions using option 1705 of the version~5.6 of the {\sc herwig} 
  generator program. This option generates  $b\bar{b}$ events with LO and
  flavor excitation diagrams, but without gluon splitting diagrams.
  We use a $b$-mass value of 3.65 $\gevcc$ and initially set the $b_s$
  lifetime equal to 1.0 ps. The $B$ hadrons, formed at the end of the
  $b$-fragmentation process, are changed into fictitious $\tilde{B}$ hadrons and 
  decayed by the {\sc herwig} spectator model fed with the
  $b_s \rightarrow l\; c\; \nu_{s} $ matrix element described in the previous
  Appendix. The decay of the $c$-hadron formed by {\sc herwig} using the
  spectator quark is modeled  with the {\sc qq} generator program~\cite{cleo}.
  Our modeling of the $b_s$ production suffers of the following approximations:
  \begin{enumerate}
  \item The $b_s \bar{b}_s$  production cross section has not the same 
        dependence on the velocity $\beta$ of the partons in the final state
        as the $b\bar{b}$ cross section~\cite{beta}. However, we study 
        jets with transverse energies larger than $15$ GeV and this alleviates
        the problem, since the corresponding partons have energies larger 
        than $20$ GeV. 
  \item The spectrum of gluon emission in the hadronization process of a spin
        0 quark differs from that of a spin 1/2 quark. However, the average 
        energy loss due to perturbative gluon emission off a spin-0 and a
        spin-1/2 particle is believed to be small~\cite{gluonem}.
 \item  Option 1705 does not evaluate the gluon splitting contribution to 
        $b_s$ production. However, we tune the $b_s$ production to reproduce
        the observed rates of tagged lepton- and away-jets. When only looking
        at yields of tagged lepton- and away-jets, Table~X of 
        Ref.~\cite{ajets} shows that events arising from flavor excitation and
        gluon splitting are quite similar. Therefore we fit the data with the
        SM~simulation implemented with the scalar quark production leaving its
        direct production and flavor excitation cross sections as additional 
        free fit parameters; in first approximation, the flavor excitation
        cross section determined by the fit also accounts for the gluon
        splitting contribution.
 \end{enumerate}

 \end{document}